\RequirePackage{ifpdf}
\ifpdf 
\documentclass[pdftex]{sigma}
\else
\documentclass{sigma}
\fi

\begin{document}

\newcommand{\sh}{{\rm sh}}
\newcommand{\ch}{{\rm ch}}
\newcommand{\De}{\Delta}
\newcommand{\de}{\delta}
\newcommand{\Z}{{\mathbb Z}}
\newcommand{\N}{{\mathbb N}}
\newcommand{\C}{{\mathbb C}}
\newcommand{\Cs}{{\mathbb C}^{*}}
\newcommand{\R}{{\mathbb R}}
\newcommand{\Q}{{\mathbb Q}}
\newcommand{\T}{{\mathbb T}}
\newcommand{\re}{{\rm Re}\, }
\newcommand{\im}{{\rm Im}\, }
\newcommand{\cW}{{\cal W}}
\newcommand{\cJ}{{\cal J}}
\newcommand{\cE}{{\cal E}}
\newcommand{\cA}{{\cal A}}
\newcommand{\cR}{{\cal R}}
\newcommand{\cP}{{\cal P}}
\newcommand{\cM}{{\cal M}}
\newcommand{\cN}{{\cal N}}
\newcommand{\cI}{{\cal I}}
\newcommand{\cMs}{{\cal M}^{*}}
\newcommand{\cB}{{\cal B}}
\newcommand{\cD}{{\cal D}}
\newcommand{\cC}{{\cal C}}
\newcommand{\cL}{{\cal L}}
\newcommand{\cF}{{\cal F}}
\newcommand{\cH}{{\cal H}}
\newcommand{\cS}{{\cal S}}
\newcommand{\cT}{{\cal T}}
\newcommand{\cU}{{\cal U}}
\newcommand{\cQ}{{\cal Q}}
\newcommand{\cV}{{\cal V}}
\newcommand{\cK}{{\cal K}}
\newcommand{\intR}{\int_{-\infty}^{\infty}}
\newcommand{\intI}{\int_{0}^{\pi/2r}}
\newcommand{\limp}{\lim_{\re x \to \infty}}
\newcommand{\limn}{\lim_{\re x \to -\infty}}
\newcommand{\limpn}{\lim_{|\re x| \to \infty}}
\newcommand{\diag}{{\rm diag}}
\newcommand{\Ln}{{\rm Ln}}
\newcommand{\Arg}{{\rm Arg}}
\newcommand{\pri}{^{\prime}}
\newcommand{\Ccs}{C_0^{\infty}(\R)}
\newcommand{\Rp}{R_{+}}
\newcommand{\dom}{D(g_0,g_1)}
\newcommand{\tP}{\tilde{\wp}}

\numberwithin{equation}{section}

\allowdisplaybreaks

\renewcommand{\thefootnote}{$\star$}

\renewcommand{\PaperNumber}{049}

\FirstPageHeading

\ShortArticleName{Hilbert--Schmidt Operators vs.  Integrable Systems III. The Heun Case}

\ArticleName{Hilbert--Schmidt Operators vs. Integrable Systems \\ of Elliptic Calogero--Moser Type III. The Heun Case\footnote{This paper is a contribution to the Proceedings of the Workshop ``Elliptic Integrable Systems, Isomonodromy Problems, and Hypergeometric Functions'' (July 21--25, 2008, MPIM, Bonn, Germany). The full collection
is available at
\href{http://www.emis.de/journals/SIGMA/Elliptic-Integrable-Systems.html}{http://www.emis.de/journals/SIGMA/Elliptic-Integrable-Systems.html}}}

\Author{Simon N.M. RUIJSENAARS~$^{\dag \ddag}$}

\AuthorNameForHeading{S.N.M. Ruijsenaars}

\Address{$^\dag$~Department of Applied Mathematics,  University of Leeds, Leeds LS2 9JT, UK}
\EmailD{\href{mailto:siru@maths.leeds.ac.uk}{siru@maths.leeds.ac.uk}}
\URLaddressD{\url{http://www.maths.leeds.ac.uk/~siru/}}

\Address{$^\ddag$~Department of Mathematical Sciences,  Loughborough University,\\
\hphantom{$^\ddag$}~Loughborough LE11 3TU, UK}

\ArticleDates{Received January 19, 2009; Published online April 21, 2009}

\Abstract{The Heun equation can be rewritten as an eigenvalue equation for an ordinary dif\/ferential operator of the form $-d^2/dx^2+V(g;x)$, where the potential is an elliptic function depending on a coupling vector $g\in\R^4$. Alternatively, this operator arises from the $BC_1$ specialization of the $BC_N$ elliptic nonrelativistic Calogero--Moser system (a.k.a.\ the Inozemtsev system). Under suitable restrictions on the elliptic periods and on $g$, we associate to this operator a self-adjoint operator $H(g)$ on the Hilbert space $\cH=L^2([0,\omega_1],dx)$, where $2\omega_1$ is the real period of $V(g;x)$. For this association and a further analysis of $H(g)$, a certain Hilbert--Schmidt operator $\cI(g)$ on $\cH$ plays a critical role. In particular, using the intimate relation of $H(g)$ and $\cI(g)$, we obtain a remarkable spectral invariance: In terms of a coupling vector $c\in\R^4$ that depends linearly on $g$, the spectrum of $H(g(c))$ is invariant under arbitrary permutations $\sigma(c)$, $\sigma\in S_4$.}


\Classification{33E05; 33E10; 46N50; 81Q05; 81Q10}

\newtheorem{lem}{Lemma}[section]
\newtheorem{theor}[lem]{Theorem}

\section{Introduction}

In the f\/irst two papers of this series (\cite{HSI, HSII}, henceforth referred to as I and II), we initiated a~detailed study of spectral properties of operators arising in the context of elliptic Calogero--Moser type systems. The operators at issue are not only the Hamiltonians def\/ining these systems, but also families of Hilbert--Schmidt (HS) operators that seem unrelated at f\/irst sight. The relation consists in the kernels of the HS operators being eigenfunctions of dif\/ferences of the Hamiltonians. As we have f\/irst detailed in our lecture notes~\cite{EIS2} and also sketched in~I, this relation can be exploited to derive novel information about the Hamiltonians, viewed as Hilbert space operators (as opposed to dif\/ferential or dif\/ference operators).

The present paper is concerned with the Heun Hamiltonian
 \begin{gather}\label{Hg}
 H(g;x) \equiv -\frac{d^2}{dx^2}+V(g;x),  \qquad g=(g_0,g_1,g_2,g_3)\in\R^4,
\\
\label{Vg}
V(g;x)\equiv  \sum_{t=0}^3g_t(g_t-1)\wp(x+\omega_t;\pi/2r,i\alpha/2),\qquad \ r,\alpha>0,
 \end{gather}
where
\begin{gather*}
\omega_0\equiv 0,\qquad \omega_1\equiv \pi/2r,\qquad \omega_2\equiv i\alpha/2,\qquad \omega_3\equiv -\omega_1-\omega_2.
\end{gather*}
(For a lucid account of elliptic and allied functions we recommend~\cite{ww}, whose conventions and notation we mostly follow.)
It can also be viewed as the $BC_1$ `nonrelativistic' elliptic Calogero--Moser Hamiltonian, the $BC_N$ case being the Inozemtsev system~\cite{inoz}. There exists an extensive lore about the eigenfunctions of $H(g)$ viewed as a dif\/ferential operator, cf.\ e.g.\ the reviews~\mbox{\cite{ronv,kuzn,maie}}. In recent years, this lore has been signif\/icantly enlarged by Takemura, whose work can be traced from his recent paper~\cite{take}.

Here, however, we are concerned with spectral properties of $H(g)$, reinterpreted as a self-adjoint operator on the Hilbert space
\begin{gather}\label{cH}
\cH \equiv L^2([0,\pi/2r],dx).
\end{gather}
A few of the properties we derive (such as the occurrence of solely discrete spectrum for suitably restricted~$g$) follow from well-known arguments in the theory of Schr\"odinger operators~\cite{rs2} (or, alternatively, from Sturm--Liouville theory and the theory of eigenfunction expansions). Our main tool for reobtaining known features and obtaining novel ones is the integral operator whose kernel is given by
 \begin{gather}\label{Psi}
 \Psi(g;x,y)\equiv w(g;x)^{1/2}\cS(g;x,y)w(g';y)^{1/2},\qquad x,y\in(0,\pi/2r).
 \end{gather}

 This kernel and the def\/inition of its ingredients can be found in~I, cf.~I~Subsection~3.2.  For ease of reference, however, we repeat the relevant def\/initions:
\begin{gather}\label{defJN}
g'\equiv J_Ng,\qquad J_N\equiv
 \frac{1}{2} \left( \begin{array}{rrrr}
1  &  1  &  -1  &  1  \\
1  &  1  &  1  &  -1  \\
-1  &  1  &  1  &  1  \\
1  &  -1  &  1  &  1
\end{array} \right) ,
\\
\label{cS}
\cS(g;x,y)\equiv \exp(-s_g\ln [R(x+y)R(x-y)]),
\\
\label{sg}
s_g\equiv \frac{1}{2} \sum_{t=0}^3g_t=\frac{1}{2}\sum_{t=0}^3g'_t.
\\
\label{w}
w(g;x)\equiv 1/c(g;x)c(g;-x),
\\
  c(g;x)\equiv R(x+i\alpha/2)^{-g_0}R(x+i\alpha/2-\pi/2r)^{-g_1}R(x)^{-g_2}R(x-\pi/2r)^{-g_3},\nonumber
\\
R(z)\equiv \prod_{l=0}^{\infty}(1-\exp[-(2l+1)r\alpha+2irz])(1-\exp[-(2l+1)r\alpha-2irz]),\qquad z\in\C.\nonumber
\end{gather}

At this point we would like to mention that a close relative of the kernel $\cS(g;x,y)$ has appeared in the literature before, a fact we only recently became aware of. Indeed, it can be tied in with the kernel that can be found on p.~124 of a monograph by Slavyanov and Lay~\cite{slla}. (It seems this kernel was f\/irst introduced in a paper by Kazakov and Slavyanov~\cite{kasl}.) More precisely, when a $\wp$-function substitution is made in the latter kernel, it can be factorized in terms of the Weierstrass $\sigma$-function; the nontrivial part is then an analytic continuation of $\cS(g;x,y)$. Both in~\cite{slla} and in Takemura's paper~\cite{take2}, this kernel is used to relate Heun functions def\/ined on dif\/ferent intervals.

The entire function $R(z)$ plays a crucial role in this paper. It satisf\/ies
\begin{gather}\label{R2}
R(z)=\exp\left( -\sum_{n=1}^{\infty}\frac{\cos(2nrz)}{n\sinh(nr\alpha)}\right),\qquad |\Im z|<\alpha/2,
\\
\label{dupl}
R(2z)=\prod_{\de =+,-}R(z+i\de \alpha/4)R(z+\pi/2r+i\de \alpha/4),
\\
\label{Rs}
R(z+i\alpha/2)=ipe^{-irz}s(z),\qquad p=2r\prod_{k=1}^{\infty}(1-e^{-2kr\alpha})^2.
\end{gather}
Here, $s(z)$ is related to the $\sigma$-function by
\begin{gather}\label{ssig}
s(z)=\exp\left(-\eta z^2r/\pi\right)\sigma(z;\pi/2r,i\alpha/2),
\end{gather}
so that
\begin{gather*}
f(z+i\alpha)=-\exp(r\alpha -2irz)f(z),\qquad f=s,R.
\end{gather*}
(For more details, see~\cite{fo}, where the functions $s$ and $R$ were introduced.)

From (\ref{Rs}) it follows that the weight function (\ref{w}) can also be written
 \begin{gather}\label{w2}
  w(g;x)=p^{2g_0+2g_1}s(x)^{2g_0}s(\pi/2r-x)^{2g_1}R(x)^{2g_2}R(\pi/2r-x)^{2g_3}.
\end{gather}
 It is positive on $(0,\pi/2r)$ for all $g\in\R^4$, but integrable on $[0,\pi/2r]$ only for $g_0,g_1>-1/2$.

 It now readily follows that for $\Psi(g;x,y)$ to belong to the kernel Hilbert space
 \begin{gather}\label{cHK}
 \cH_K\equiv L^2\left([0,\pi/2r]^2,dxdy\right),
 \end{gather}
  the necessary and suf\/f\/icient condition is that $g$ belong to the parameter set
 \begin{gather}\label{Pi}
 \Pi \equiv \left\{ g\in\R^4\mid g_0,g_1,g_0',g_1'>-1/2\right\}.
 \end{gather}
 Hence it is clear that the integral operator
 \begin{gather}\label{cI}
 (\cI(g) f)(x)\equiv \intI dy\Psi(g;x,y)f(y),\qquad f\in\cH,\qquad
g\in\Pi,
\end{gather}
 is well def\/ined and HS.

 It is not obvious, but true that its kernel is a zero-eigenvalue eigenfunction of a dif\/ference of Heun Hamiltonians:
\begin{gather}\label{eiB}
(H(g;x)-H(g';y))\Psi(g;x,y)=0,\qquad x,y\in(0,\pi/2r).
\end{gather}
(Indeed, this is a corollary of Proposition~3.4 in I, cf.~I(3.115).) At face value, this eigenfunction identity seems to have no bearing on the Hilbert space eigenfunctions of the two (generically) distinct Hamiltonians. A principal result of this paper is, however, that
for $g$ in the restricted parameter set
\begin{gather}\label{Pir}
\Pi_r\equiv \{ g\in\Pi \mid s_g>0\},
\end{gather}
the Hilbert space eigenvector ONBs (orthonormal bases) of $H(g)$ and $H(g')$ are given by functions $e_n(g;x)$ and $e_n(g';x)$, resp., which yield the singular value decomposition of $\cI(g)$ (cf. \cite{rs1}):
\begin{gather}\label{sval}
\cI(g)=\sum_{n=0}^{\infty}\nu_n(g) e_n(g)\otimes e_n(g'),\qquad \nu_0\ge \nu_1\ge \cdots > 0,
\qquad \sum_{n=0}^{\infty}\nu_n^2<\infty,
\\
(e_m,e_n)=\de_{mn},\qquad m,n\in\N.\nonumber
\end{gather}
(All singular values are positive, since $\cI(g)$ has trivial kernel and dense range for $g\in\Pi_r$, cf.~(\ref{KerI}). To appreciate the restriction on $s_g$, note that $\cI(g)$ has rank 1 for $s_g=0$, cf.~(\ref{cS}).)

This result entails in particular that $H(g)$ and $H(g')$ have solely discrete spectrum, the two spectra being equal.
To be sure, in the case at hand it follows from well-known Schr\"odinger operator theory that $H(g)$ can be reinterpreted as a self-adjoint operator that has an ONB of eigenvectors, provided $g$ is suitably restricted. The main novelty is that these eigenvectors are just those occurring for $\cI(g)$, and that this f\/ixes the self-adjoint extensions of the Schr\"odinger operator $H(g)$ in case it is not essentially self-adjoint on $C_0^{\infty}((0,\pi/2r))$ (namely, for $g_0$ and/or $g_1$ in $(-1/2,3/2)$). As a bonus, this yields a hidden spectral invariance under permutations of four new parameters, which implies spectral consequences for special cases that are quite startling. Last but not least, the state of af\/fairs for the present case gives rise to a simple paradigm for the `relativistic' (analytic dif\/ference operator) Heun case, where the usual self-adjoint extension theory is of no help and no previous spectral results are known, cf.\ the fourth paper in this series~\cite{HSIV}.

We proceed to sketch the organization and results of this paper in more detail.  We begin Section~\ref{section2} by deriving features of functions in the range of $\cI(g)$, assuming $g\in\Pi_r$. Then we show that $\cI(g),g\in\Pi_r$, has dense range and trivial kernel.  These crucial features readily follow from Lemma~\ref{lemma2.1}, which asserts that the integral operator with kernel $\cS(g;x,y)$ has kernel $\{ 0\}$ on $L^1([0,\pi/2r])$ whenever $s_g$ is positive.
The characteristics of the range of $\cI(g)$ are now used as a guide for a choice of domain for $H(g)$, namely the dense subspace $D(g_0,g_1)$ given by (\ref{D01}). Lemma~\ref{lemma2.2} then asserts that $H(g)$ is symmetric on $D(g_0,g_1)$ for all $g$ in the parameter space
\begin{gather}\label{tPi}
\tilde{\Pi}\equiv \left\{ g\in\R^4 \mid g_0,g_1>-1/2\right\}.
\end{gather}
  The proofs of these lemmas involve technicalities that are not enlightening, so we have relegated them to Appendix~\ref{appendix-A}.

Next, using the key identity (\ref{eiB}), we show in Lemma~\ref{lemma2.3} that for all $g\in\tilde{\Pi}$ the subspace $\dom$ is a core (domain of essential self-adjointness~\cite{rs2}). This prepares us for studying the self-adjoint closure (again denoted $H(g)$) by using $\cI(g)$. Only a few simple steps are then needed to reach the conclusion that the above ONB $\{e_n(g)\}$ may be chosen so that it is an eigenvector ONB for $H(g)$. Moreover, since the eigenvectors belong to $\dom$, it easily follows from the Frobenius method that the spectrum of $H(g)$ is simple,  cf. Theorem~\ref{theorem2.4}.

Whenever the potential $V(g;x)$ is bounded below, it is obvious that $H(g)$ is also bounded below. For $g_0$ and/or $g_1$ in $ (0,1)$, however, $V(g;x)$ is not bounded below, so an extra argument is needed to show that for the pertinent self-adjoint extensions of (\ref{Hg}) semi-boundedness is preserved. We prove this by a comparison argument involving the trigonometric $BC_1$ Calogero--Moser Hamiltonian
\begin{gather}
H_t(g_0,g_1;x)\equiv -\frac{d^2}{dx^2}+V_t(g_0,g_1;x),\nonumber\\
V_t(g_0,g_1;x)\equiv \frac{r^2g_0(g_0-1)}{\sin^2rx}+\frac{r^2g_1(g_1-1)}{\cos^2rx}.\label{Ht}
\end{gather}
The point is that this operator is well def\/ined and essentially self-adjoint on $\dom$, and that it has an ONB of eigenvectors in $\dom$  involving the well-known Jacobi polynomials, cf.~Theorem~\ref{theorem2.5}.

We would like to add that the trigonometric Hamiltonian (\ref{Ht}) plays a key role in Takemura's paper~\cite{TakeHeun}, where he studies the Heun Hamiltonian $H(g;x)$ (\ref{Hg}) with $g_0,g_1\ge 1$ from a Hilbert space point of view. His strategy is based on a perturbation expansion in the parameter $p=\exp(-\alpha r)$. (In this connection it should be pointed out that $H(g;x)$ with $g_2=g_3=0$ reduces to $H_t(g_0,g_1;x)$ for $\alpha\to\infty$.)

The main result of Theorem~\ref{theorem2.6} is that for $g\in\Pi_r$ the spectra of $H(g)$ and $H(g')$ coincide. Unfortunately, we have not been able to show that the singular values in (\ref{sval}) are distinct for all $g\in\Pi_r$, a feature that seems plausible to us. If this is indeed the case, then it would follow that $e_n(g)$ is the $H(g)$-eigenvector with eigenvalue $E_n(g)$ in the obvious eigenvalue list
\begin{gather*}
E_0(g)<E_1(g)<E_2(g)<\cdots,\qquad g\in\Pi_r.
\end{gather*}

In Section~\ref{section3} we f\/ix $g\in\Pi_r$ and address the question for which $\hat{g}\in\tilde{\Pi}$ the spectrum of $H(\hat{g})$ coincides with that of $H(g)$. From Theorem~\ref{theorem2.6} we know this is true for $\hat{g}=g'$, but there are a few more involutory $g$-transformations for which this is valid.  We determine the group $G$ generated by these involutions in Theorem~\ref{theorem3.1}. It leaves the parameter set
\begin{gather}\label{PiG}
\Pi_G\equiv \{ g\in\Pi_r\mid g_0+g_1+2>g_2+g_3\}
\end{gather}
invariant, and is isomorphic to the permutation group $S_4$.
(For ease of reference, we collect the four parameter sets (\ref{PiG}), (\ref{Pir}), (\ref{Pi}) and (\ref{tPi}) in a chain
\begin{gather*}
\Pi_G\subset \Pi_r\subset \Pi\subset\tilde{\Pi},
\end{gather*}
and note that each inclusion is proper.) Indeed, def\/ining new couplings
\begin{gather*}
c_0\equiv g_0+g_3-1,\qquad  c_1\equiv g_1+g_2-1,\qquad c_2\equiv g_1-g_2,\qquad  c_3\equiv g_0-g_3,
\end{gather*}
the spectrum of
\begin{gather*}
\hat{H}(c)\equiv H(g(c))
\end{gather*}
 is invariant under arbitrary permutations of $c_0$, $c_1$, $c_2$, $c_3$.

 In Section~\ref{section4} we consider various special $g$-choices. For the sixteen $g$-values yielding \mbox{$V(g;x){=}0$}, all of the relevant quantities can be determined in great detail. A striking consequence of our spectral invariance results is that there exist quite a few $g$-choices  for which $V(g;x)\ne 0$, yet the spectrum of $H(g)$ equals that  for one of the cases where $V(g;x)=0$. We also consider the case $s_g=0$, for which $\cI(g)$ has rank~1. In particular, using results from~I and~\cite{cadiz}, we verify that $\cI(g)$ connects the ground state of $H(g')$ with the one of $H(g)$. Finally, we add some observations on the case where the $\pi/r$-periodic potential $V(g;x)$ has no poles for real $x$, a well-known setting that is for instance studied in considerable detail in~\cite{arsc}.

The paper is concluded with Appendix~\ref{appendix-A}, in which we present the proofs of Lemmas~\ref{lemma2.1} and~\ref{lemma2.2}.

\section[The eigenvector ONBs of $H(g)$ and $H(g')]{The eigenvector ONBs of $\boldsymbol{H(g)}$ and $\boldsymbol{H(g')}$}\label{section2}

Unless explicitly mentioned otherwise, we assume from now on that the coupling vector $g$ belongs to the restricted parameter set~$\Pi_r$~(\ref{Pir}). This entails in particular that the integral operator~$\cI(g)$~(\ref{cI}) is a well-def\/ined HS operator on the Hilbert space~$\cH$~(\ref{cH}). We begin this section by obtaining features of functions in the range of $\cI(g)$.

Consider f\/irst for $f\in\cH$ and $x\in[0,\pi/2r]$ the function
\begin{gather}
h(x)     \equiv    \int_0^{\pi/2r}dy\cS(g;x,y)w(g';y)^{1/2}f(y)
\nonumber\\
\phantom{h(x)}{}   =       \int_0^{\pi/2r}dy\exp\left( 2s_g\sum_{n=1}^{\infty}\frac{\cos(2nrx)\cos(2nry)}{n\sinh(nr\alpha)}\right) w(g';y)^{1/2}f(y),\label{hdef}
  \end{gather}
  where we used (\ref{R2}). Since we have $w(g';\cdot)^{1/2}\in\cH$, we deduce
  from the Schwarz inequality
  \begin{gather*}
  w(g';y)^{1/2}f(y)\in L^1([0,\pi/2r],dy).
  \end{gather*}
  It readily follows that $h(x)$ extends to a function $h(z)$, $\Re z=x$, that is analytic in the strip $|\Im z|<\alpha/2$. Also, $h(z)$ is $\pi/r$-periodic and even. Now $R(z)$ is an entire, $\pi/r$-periodic and even function without zeros for $|\Im z|<\alpha/2$, so the function
  \begin{gather}\label{ana}
  s(x)^{-g_0}s(\pi/2r-x)^{-g_1}(\cI(g) f)(x)=R(x)^{g_2}R(\pi/2r-x)^{g_3}h(x),\qquad x\in (0,\pi/2r),
  \end{gather}
  extends to a function that is analytic for $|\Im z|<\alpha/2$, $\pi/r$-periodic and even.

  Next, we point out the relations
  \begin{gather}
  g\in\Pi_r \ \Leftrightarrow \ g'\in\Pi_r, \nonumber
  \\
  \label{cIad}
  \cI(g)^{*}=\cI(g'),
  \end{gather}
  which are easily verif\/ied. As a consequence, the orthocomplement of the range of $\cI(g)$ equals the kernel of $\cI(g')$. From (\ref{hdef}) it is immediate that $\cI(g), g\in\Pi$, has rank 1 when $s_g$ vanishes.  By contrast, for $s_g>0$ the integral operator $\cI(g)$ has dense range and trivial kernel:
\begin{gather}\label{KerI}
{\rm Ker}( \cI(g))=\{ 0\},\qquad {\rm Ran}(\cI(g))^{-}=\cH,\qquad \forall \, g\in\Pi_r.
\end{gather}
This is clear from the above and the following lemma, which is proved in Appendix~\ref{appendix-A}.

\begin{lem}\label{lemma2.1} Letting $s_g>0$, suppose $\phi(y) \in L^1([0,\pi/2r],dy)$ satisfies
\begin{gather*}
 \int_0^{\pi/2r}dy\cS(g;x,y)\phi(y)=0.
\end{gather*}
Then $\phi =0$.
\end{lem}

To proceed, we note that the kernel function $\Psi(g;x,y)$ is real-valued. From (\ref{KerI}), (\ref{cIad}) and the HS property we then deduce that for a f\/ixed $g\in\Pi_r$, there exist two ONBs $\{ e_n(g')\}_{n\in\N}$ and $\{ e_n(g)\}_{n\in\N}$ of real-valued functions such that
\begin{gather}\label{Psirep}
\Psi(g;x,y)=\sum_{n=0}^{\infty}\nu_n(g)e_n(g;x)e_n(g';y),\qquad \nu_0\ge\nu_1\ge \cdots >0,
\end{gather}
with $\{ \nu_n\}\in l^2(\N)$. (The choice of the f\/irst ONB uniquely determines the second one; later on, we will return to the specif\/ic choice we make.) Moreover, since $e_n(g)$ belongs to the range of~$\cI(g)$, the above analysis implies that it is of the form
\begin{gather}\label{ed}
e_n(g;x)=w(g;x)^{1/2}d_n(g;x),
\end{gather}
where $d_n(g;x)$ extends to a function that is analytic in $|\Im z|<\alpha/2$, even and $\pi/r$-periodic.

We are now prepared to deal with the Heun Hamiltonian $H(g)$. We choose as initial domain for $H(g)$ the space
\begin{gather}\label{D01}
D(g_0,g_1)=s(x)^{g_0}s(\pi/2r-x)^{g_1}\cD,\qquad g_0,g_1>-1/2,
\end{gather}
where
\begin{gather}\label{cD}
\cD \equiv \left\{ f\in C^2([0,\pi/2r])\mid f'(0)=f'(\pi/2r)=0\right\}.
\end{gather}
This domain choice ensures that the ONB functions $e_n(g;x)$ belong to the domain. Indeed,
from the paragraph containing (\ref{ana}) we see that the function on the r.h.s.\ of (\ref{ana})  belongs to $\cD$, so that we have more generally
\begin{gather}\label{Ran}
{\rm Ran}\, (\cI(g))\subset D(g_0,g_1).
\end{gather}

From $e_n(g)\in D(g_0,g_1)$ it already follows that $D(g_0,g_1)$ is dense. But this feature can also be derived quite easily without invoking $\cI(g)$. Indeed, $\cD$ contains the functions $\cos(2nrx)$, $n\in\N$, which are well known to yield an orthogonal base of $\cH$. Therefore, the assumption $f\perp D(g_0,g_1)$ entails
\begin{gather*}
\int_0^{\pi/2r}dxs(x)^{g_0}s(\pi/2r-x)^{g_1}\cos(2nrx)f(x)=0,\qquad \forall \, n\in\N.
\end{gather*}
From the orthogonal base property we now deduce $f(x)=0$, so that $D(g_0,g_1)$ is dense.

We are now prepared for our second lemma, whose proof is relegated to Appendix~\ref{appendix-A}.

\begin{lem}\label{lemma2.2}
Assume $g$ belongs to $\tilde{\Pi}$ \eqref{tPi}.
Then $H(g)$ \eqref{Hg} is well defined and symmetric on the domain $D(g_0,g_1)$ given by \eqref{D01}, \eqref{cD}.
\end{lem}

The proof of this lemma does not involve the HS operator $\cI(g)$. In the next lemma we do use $\cI(g)$ to obtain a short proof of essential self-adjointness on $D(g_0,g_1)$.

\begin{lem}\label{lemma2.3}
Assuming $g\in\tilde{\Pi}$,  the domain $D(g_0,g_1)$ is a core for $H(g)$.
\end{lem}

\begin{proof}
We begin by noting that the potentials $\wp(x+i\alpha/2)$ and $\wp(x-\pi/2r-i\alpha/2)$ yield bounded self-adjoint operators, so that they do not inf\/luence domain issues. Therefore, we need only prove the core property for one value of $g_2$ and $g_3$. We choose $g_2=g_3=1$ (say), the point being that this entails $g\in\Pi_r$. (This is easily checked, cf.~(\ref{defJN}), (\ref{Pi}) and (\ref{Pir}).) Hence we can now use $\cI(g)$.

Assume $\phi\in\cH$ satisf\/ies
\begin{gather*}
(\phi,(H(g)+i)\psi)=0,\qquad \forall \, \psi\in D(g_0,g_1).
\end{gather*}
By virtue of (\ref{Ran}), this implies in particular
\begin{gather*}
(\phi,(H(g)+i)\cI(g) f)=0,\qquad \forall \, f\in\cH.
\end{gather*}
Thanks to (\ref{eiB}) and (\ref{cIad}), this inner product can be rewritten as
\begin{gather*}
((H(g')-i)\cI(g')\phi,f)=0,\qquad \forall \, f\in\cH.
\end{gather*}
This implies that the vector $\cI(g')\phi$, which belongs to $D(g_0^{'},g_1^{'})$ by (\ref{Ran}), is an eigenvector with eigenvalue $i$ for the symmetric operator $H(g')$.  Thus we must have $\cI(g')\phi=0$. By (\ref{KerI}), this entails $\phi=0$. Therefore ${\rm Ran}\, (H(g)+i)$ is dense in $\cH$. Likewise, ${\rm Ran}\, (H(g)-i)$ is dense in $\cH$.  Hence the core property results.
\end{proof}

We would like to point out that the dense subspace
\begin{gather}\label{cC}
\cC\equiv C_0^{\infty}((0,\omega_1))\subset D(g_0,g_1),
\end{gather}
is already a core for $H(g)$ whenever $g_0,g_1> 3/2$. This is basically known, but for completeness we sketch a proof. Choosing at f\/irst $g\in\R^4$, consider the dif\/ferential operator $H(g)$ restricted to $\cC$. This is obviously a well-def\/ined symmetric operator on $\cH$, and it is easy to verify that the domain of its closure contains the space
\begin{gather*}
\cC'    \equiv     \big\{ \psi\in C^1([0,\omega_1])\mid \psi(0)=\psi(\omega_1)=\psi'(0)=\psi'(\omega_1)=0,
\nonumber \\
\phantom{\cC'    \equiv     \big\{}{}  \psi'(x)\in C^1((0,\omega_1)), \psi''(x), x^{-2}\psi(x),(\omega_1-x)^{-2}\psi(x)\in\cH\big\}.
\end{gather*}
Now assume $\phi$ is orthogonal to $ (H(g)+i)\cC$. Then $\phi$ is a weak solution to the ODE $(H(g)-i)\phi=0$ on $(0,\omega_1)$, and by hypo-ellipticity it is a classical solution.

Choosing next $g_0,g_1> 3/2$, it follows from the Frobenius theory that near 0 we have
\begin{gather*}
\phi(x)= x^{g_0}h_0(x),
\end{gather*}
with $h_0(x)$ analytic at $x= 0$. (Indeed, $x^{1-g_0}$ is not in $L^2$ near 0.) Likewise, near $x=\omega_1$,
\begin{gather*}
 \phi(x)= (\omega_1-x)^{g_1}h_1(x),
 \end{gather*}
 with $h_1(x)$ analytic at $x=\omega_1$. But then we deduce that $\phi$ has all of the properties def\/ining $\cC'$, and since the closure of $H(g)$ is symmetric on $\cC'$, it cannot have an eigenvector with nonreal eigenvalue. Thus $\phi=0$ and the core property follows.

After this digression we proceed to obtain our f\/irst main result, for which we again need to require $g\in\Pi_r$. We denote the domain of the self-adjoint closure of $H(g)$ by $D(g)$, but keep the notation $H(g)$ for the closure. We now use the key identity (\ref{eiB}) as in the proof of Lemma~\ref{lemma2.3} to deduce
\begin{gather}\label{key2}
(\phi,H(g)\cI(g)\psi)=(H(g')\cI(g')\phi,\psi),\qquad \forall \, \phi,\psi\in\cH.
\end{gather}
Choosing $\lambda\in\C$ with $\Im \lambda >0$, this entails in particular
\begin{gather}\label{ipH}
(\phi,(H(g)+\lambda)\cI(g)\psi)=((H(g')+\overline{\lambda})\cI(g')\phi,\psi),\qquad \forall \, \phi\in D(g),\qquad \forall \, \psi\in D(g').
\end{gather}
Now for $\phi\in D(g)$ and $\psi\in D(g')$ there exist $e,f\in\cH$ such that
\begin{gather*}
\phi=(H(g)+\overline{\lambda})^{-1}e,\qquad \psi=(H(g')+\lambda)^{-1}f,
\end{gather*}
so (\ref{ipH}) implies
\begin{gather*}
(e,\cI(g)(H(g')+\lambda)^{-1}f)=(\cI(g')(H(g)+\overline{\lambda})^{-1}e,f),\qquad \forall \, e,f\in\cH.
\end{gather*}
Hence we obtain the relation
\begin{gather*}
\cI(g)(H(g')+\lambda)^{-1}=(H(g)+\lambda)^{-1}\cI(g).
\end{gather*}
From this we readily deduce
\begin{gather*}
[(H(g)+\lambda)^{-1},T(g)]=0,
\end{gather*}
where
\begin{gather}\label{T}
T(g)\equiv \cI(g)\cI(g)^{*}=\sum_{n=0}^{\infty}\nu_n^2(g)e_n(g)\otimes e_n(g).
\end{gather}

The upshot is that the self-adjoint trace class operator $T(g)$  commutes with the $H(g)$-resolvent. Since $T(g)$ has trivial kernel by~(\ref{KerI}), all of its eigenspaces are f\/inite-dimensional, and the $H(g)$-resolvent leaves these f\/inite-dimensional spaces  invariant. Thus we can choose the eigenvectors of $T(g)$ to be eigenvectors of the $H(g)$-resolvent, hence of $H(g)$ as well. We now summarize and extend these f\/indings.

\begin{theor}\label{theorem2.4}
Let $g\in\Pi_r$. Then the positive trace class operator $T(g)$~\eqref{T} has an ONB $\{ e_n(g)\}_{n\in\N}$, of eigenvectors that are also eigenvectors of $H(g)$. The functions $e_n(g;x)$ are real-valued and belong to $\dom$. They are of the form \eqref{ed}, where $d_n(g;x)$ extends to a function that is analytic
in $|\Im z|<\alpha/2$, even and $\pi/r$-periodic.
Moreover, we have
\begin{gather}\label{dnnz}
d_n(g;0)\ne 0,\qquad d_n(g;\pi/2r)\ne 0,
\end{gather}
and the spectrum of $H(g)$ is nondegenerate.
\end{theor}

\begin{proof} We have just proved the f\/irst assertion. The second and third one have been shown earlier too, cf.~the paragraphs containing (\ref{ed}) and (\ref{cD}). To prove the last one, we f\/irst note that it follows from $e_n(g)\in\dom$ that $e_n(g;x)$ is a classical solution to the ODE $H(g)e_n=\lambda_ne_n$ for a certain $\lambda_n\in\R$. From the Frobenius method it then follows that $d_n(g;x)$ satisf\/ies (\ref{dnnz}). Moreover, any solution to the same ODE that is linearly independent of $e_n(g;x)$ does not have the features encoded in (\ref{ed}), so the eigenvalue $\lambda_n$ is simple.
\end{proof}

 Quite likely, for $g\in\tilde{\Pi}\setminus \Pi_r$ the $H(g)$-eigenvectors still belong to $\dom$ and have the same features as for $g\in\Pi_r$. (If so, it follows as before that $\sigma(H(g))$ is simple.) However, we cannot prove this via the above arguments, since they hinge on exploiting the HS operator $\cI(g)$ with $g$ restricted to $\Pi_r$.

It is a corollary of Theorem~\ref{theorem2.4} that there exist distinct real numbers $\lambda_n(g)$ such that
\begin{gather}\label{Hlam}
H(g)e_n(g)=\lambda_n(g)e_n(g), \qquad n\in\N,\qquad g\in\Pi_r.
\end{gather}
At this stage, however, we have no other information on the $H(g)$-eigenvalues.
In the following theorem, we obtain further results on the $H(g)$-spectrum that do not involve $\cI(g)$, and which are valid for all $g\in\tilde{\Pi}$. As a preparation, note that $H(g)$ is obviously bounded below for $g_0,g_1\ge 1$. (Indeed, this entails that $V(g;x)$ is bounded below, cf.~(\ref{Vg}).) For $g_0$ and/or $g_1$ in $(0,1)$, however, $V(g;x)$ diverges to $-\infty$ as $x\downarrow 0$ and/or $x\uparrow \pi/2r$. Even so,  our next theorem implies that $H(g)$ is still bounded below. Its proof is based on a comparison argument involving the trigonometric $BC_1$ Calogero--Moser Hamiltonian~(\ref{Ht}).

\begin{theor}\label{theorem2.5}
For all $g\in\tilde{\Pi}$, the spectrum of the operator $H(g)$ is a discrete set that is bounded below. Moreover, each eigenvalue has finite multiplicity.
\end{theor}

\begin{proof}  The trigonometric and elliptic dif\/ferential operators (\ref{Ht}) and (\ref{Hg}) dif\/fer by a real-valued potential
\begin{gather*}
V_d(g;x)\equiv V_t(g_0,g_1;x)-V(g;x),
\end{gather*}
 that is bounded on $[0,\pi/2r]$. Hence we can associate to (\ref{Ht}) the operator on $\dom$ given by
\begin{gather}\label{Vd}
H_t(g_0,g_1)\equiv H(g)+V_d(g;\cdot).
\end{gather}
This operator is essentially self-adjoint on $\dom$, since $H(g)$ is e.s.a.~on $\dom$ by virtue of Lemma~\ref{lemma2.3} and the multiplication operator $V_d(g;\cdot)$ is bounded and self-adjoint.
We now introduce the trigonometric weight function
\begin{gather*}
w_t(g_0,g_1;x)\equiv (\sin rx)^{2g_0}(\cos rx)^{2g_1},
\end{gather*}
and observe that we may rewrite $\dom$ as
\begin{gather*}
\dom =w_t(g_0,g_1;x)^{1/2}\cD.
\end{gather*}
(Indeed, $\cD$ (\ref{cD}) is closed under products and we have
\begin{gather*}
\left(\frac{s(x)}{\sin rx}\right)^{\lambda},\ \ \left(\frac{s(\pi/2r-x)}{\cos rx}\right)^{\mu}\in\cD,\qquad \lambda,\mu\in\R,
\end{gather*}
as is easily checked.)

Next, we calculate the similarity-transformed dif\/ferential operator
\begin{gather*}
A_t(g_0,g_1;x)     \equiv     w_t(g_0,g_1;x)^{-1/2}H_t(g_0,g_1;x)w_t(g_0,g_1;x)^{1/2}
\nonumber \\
 \phantom{A_t(g_0,g_1;x)}{}  =     -\frac{d^2}{dx^2}-2r(g_0\cot rx -g_1\tan rx)\frac{d}{dx}+r^2(g_0+g_1)^2.
\end{gather*}
This yields an essentially self-adjoint operator $A_t(g_0,g_1)$ on the dense subspace $\cD$ (given by~(\ref{cD})) of the Hilbert space
\begin{gather*}
\cH_t\equiv L^2([0,\pi/2r],w_t(g_0,g_1;x)dx).
\end{gather*}
The crux is now that $A_t(g_0,g_1)$ has an orthogonal base of eigenvectors belonging to $\cD$ with nonnegative eigenvalues, explicitly given by
\begin{gather}\label{dmt}
d_{m,t}(g_0,g_1;x)= {}_2F_1(-m,g_0+g_1+m;g_0+1/2;\sin^2rx),\qquad m\in\N,
\\
\label{teig}
A_t(g_0,g_1)d_{m,t}(g_0,g_1)=r^2(g_0+g_1+2m)^2d_{m,t}(g_0,g_1),\qquad m\in\N.
\end{gather}
Indeed, the r.h.s.\ of (\ref{dmt}) yields polynomials of degree $m$ in $\sin^2rx$, and these polynomials amount to the Jacobi polynomials
\begin{gather*}
P_m^{(\alpha,\beta)}(y),\qquad \alpha\equiv g_0-1/2,\qquad \beta\equiv g_1-1/2,\qquad y\equiv \cos 2rx,\qquad  m\in\N,
\end{gather*}
which form an orthogonal base in
\begin{gather*}
\cH_P\equiv L^2\big([-1,1],(1-y)^{\alpha}(1+y)^{\beta}dy\big),
\end{gather*}
see for instance~\cite{kosw}.

From the above it follows that the spectrum of $H_t(g_0,g_1)$, $g_0,g_1>-1/2$, is bounded below, so by (\ref{Vd}) the same is true for $H(g)$, $g\in\tilde{\Pi}$. Furthermore, it is clear from the eigenvalues (\ref{teig}) that the resolvent of $H_t(g_0,g_1)$ is compact. By the second resolvent identity this also holds true for the resolvent of $H(g)$, so the theorem follows.
\end{proof}

This theorem also reveals the special character of the (at most four) self-adjoint extensions of $H(g)$ restricted to $\cC$ (\ref{cC}) that arise via $\cI(g)$: They are the extensions whose trigonometric limits yield the (at most four) self-adjoint extensions of $H_t(g_0,g_1)$ restricted to $\cC$ that are associated with the Jacobi polynomials. Of course, this is particularly clear when $V(g;x)$ and $V_t(g_0,g_1;x)$ vanish identically, cf.~Section~\ref{section4}.

For the remainder of this section we assume $g\in\Pi_r$. It follows from Theorems~\ref{theorem2.4} and~\ref{theorem2.5} that $H(g)$ has nondegenerate eigenvalues
\begin{gather*}
E_0(g)<E_1(g)<E_2(g)<\cdots.
\end{gather*}
Also, since $H(g)$ and $H_t(g_0,g_1)$ dif\/fer by a bounded self-adjoint operator (cf.~(\ref{Vd})), it follows from (\ref{teig}) that the eigenvalues have asymptotics
\begin{gather*}
E_m(g)\sim 4r^2m^2,\qquad m\to\infty.
\end{gather*}
By Theorem~\ref{theorem2.4} the associated eigenvectors $u_m(g)$ are also eigenvectors of $T(g)$, and they are just the scalar multiples of $e_n(g)$ for some $n\in\N$. Hence we may and will f\/ix  the ordering choice of the $T(g)$-eigenvectors for a given degenerate eigenvalue (if any occur), attained for example for $\nu_j=\nu_{j+1}=\cdots =\nu_{j+k}$, by requiring that the associated $H(g)$-eigenvalues satisfy $\lambda_j<\lambda_{j+1}<\cdots <\lambda_{j+k}$ . With this ordering of the $e_n(g)$ understood from now on, there is a~uniquely determined permutation $\tau_g$ of the nonnegative integers such that
\begin{gather}\label{perm}
\lambda_{\tau_g(m)}(g)=E_m(g),\qquad m\in\N,
\end{gather}
cf.~(\ref{Hlam}).

Next, we observe that we still have not completely f\/ixed the choice of the basis vectors $e_n(g)$. Indeed, we have required real-valuedness of $e_n(g;x)$, but this leaves a sign undetermined. We may and will f\/ix this ambiguity by requiring
\begin{gather*}
d_n(g;0)>0,\qquad \forall \, n\in\N,
\end{gather*}
cf.~(\ref{dnnz}).

At this point the alert reader might (or rather should) object that the sign choice just detailed may be in conf\/lict with the decomposition (\ref{Psirep}). Indeed, once the ONB-functions $e_n(g;x)$ have been f\/ixed, the ONB-functions $e_n(g';x)$ are uniquely determined, and the signs of the latter might dif\/fer from the above sign convention.

It follows from the last theorem of this section that this contingency does not arise. Its main result is, however, that $H(g)$ and $H(g')$ have the same spectrum.

\begin{theor}\label{theorem2.6}
The integral operator $\cI(g)$ is real-analytic in $g$ on $\Pi_r$ in the Hilbert--Schmidt norm. With the above sign convention for $e_n(g;x)$ understood, it has singular value decomposition \eqref{sval}. Furthermore, we have an equality of spectra
\begin{gather*}
\sigma(H(g))=\sigma(H(g')),\qquad g\in\Pi_r.
\end{gather*}
\end{theor}

\begin{proof}  Analyticity of $\cI(g)$ in $g$ with respect to the HS norm amounts to analyticity of its ker\-nel $\Psi(g;x,y)$ in $g$ in the strong topology of the Hilbert space $\cH_K$ (\ref{cHK}). From the explicit for\-mula~(\ref{Psi}) it is readily verif\/ied that the kernel is strongly analytic in $g$ in a complex neighborhood of $\Pi_r$, so the f\/irst assertion follows.

Next, we f\/ix $g\in\Pi_r$ and note that the vectors
\begin{gather*}
b_m(g;x)\equiv e_{\tau_g(m)}(g;x),\qquad m\in\N,
\end{gather*}
yield an ONB with $H(g)$-eigenvalues $E_m(g)$, cf.~(\ref{perm}). Recalling (\ref{key2}) and (\ref{cIad}), we now deduce
\begin{gather*}
H(g')\cI(g')b_m(g)   =  \cI(g')H(g)b_m(g)
 =  E_m(g)\cI(g')b_m(g).
\end{gather*}
In view of (\ref{KerI}), this entails that the vectors $\cI(g')b_m(g)$ are eigenvectors of $H(g')$ with eigenvalues $E_m(g)$, and that they are total in $\cH$. Hence,
\begin{gather*}
E_m(g)=E_m(g'),\qquad \forall \, m\in\N,
\end{gather*}
and so the last assertion follows. Moreover, it follows that we have
\begin{gather*}
\tau_g=\tau_{g'},
\end{gather*}
and
\begin{gather*}
\cI(g')b_m(g)=\mu_m(g')b_m(g'),\qquad \mu_m(g')\in\R^{*}.
\end{gather*}

To prove the theorem, it remains to show that all of the numbers $\mu_m(g)$ are positive. Now they do not vanish and are continuous on $\Pi_r$, since their squares yield eigenvalues of the self-adjoint trace class operator $T(g)$, which (by the f\/irst assertion) is continuous in $g$ on $\Pi_r$ in the trace norm topology. As $\Pi_r$ is a connected set, it suf\/f\/ices to check positivity for one $g$ in $\Pi_r$.  For $g=(0,0,1,1)$ we have $g'=g$ and $\cI(g)$ is self-adjoint. Thus the numbers $\mu_m(g)$ are eigenvalues of $\cI(g)$. These are easily determined explicitly, and they are all positive, cf.~(\ref{nun}).
\end{proof}

For all of the special $g\in\Pi_r$ where we have calculated the numbers $\mu_m(g)$ explicitly (cf.~Section~\ref{section4}), they are distinct and decreasing in $m$:
\begin{gather*}
\mu_0(g)>\mu_1(g)>\mu_2(g)>\cdots >0.
\end{gather*}
This entails in particular that the permutation $\tau_g$ is the identity permutation. It is an open problem whether these features hold true on all of $\Pi_r$.

\section[A hidden $S_4$ symmetry]{A hidden $\boldsymbol{S_4}$ symmetry}\label{section3}

In this section our purpose is to study, for a given coupling vector $g\in \Pi_r$, the couplings $\hat{g}$ for which the spectra of $H(g)$ and $H(\hat{g})$ are equal.
To this end we note that we can establish equality of spectra by exploiting the core property of $\dom$, cf.~Lemma~\ref{lemma2.3}. First, since the action of the dif\/ferential operator $H(g)$ on the domain $\dom$ is invariant under taking $g_t\to 1-g_t$, and the domain does not depend on~$g_2$ and~$g_3$, we get equality of spectra (and even equality of operators) under the transformations
\begin{gather}\label{triv}
g_2\to 1-g_2,\qquad g_3\to 1-g_3.
\end{gather}
(A caveat is in order: these maps can take $g$ out of $\Pi_r$.) Less trivially, consider the permutation
\begin{gather}\label{refl}
(g_0,g_1,g_2,g_3)\to (g_1,g_0,g_3,g_2).
\end{gather}
Def\/ining the `mirror' unitary
\begin{gather}\label{mir}
(Mf)(x)\equiv f(\pi/2r-x),\qquad x\in[0,\pi/2r],\qquad f\in\cH,
\end{gather}
we clearly get
\begin{gather*}
M\dom =D(g_1,g_0).
\end{gather*}
This implies
\begin{gather*}
MH(g_0,g_1,g_2,g_3)M=H(g_1,g_0,g_3,g_2),
\end{gather*}
so that (\ref{refl}) yields operators with equal spectra; in this case the two operators are dif\/ferent for generic $g$.

Next, we note that the matrix $J_N$ (\ref{defJN}) can be viewed as the ref\/lection with respect to the unit vector $(1,-1,1,-1)/2$:
\begin{gather*}
J_N={\bf 1}_4-2u\otimes u,\qquad u\equiv (1,-1,1,-1)/2.
\end{gather*}
Thus the self-dual couplings are given by
\begin{gather*}
g'=g \ \Leftrightarrow \ g_0+g_2=g_1+g_3,
\end{gather*}
and on this 3-dimensional submanifold of $\Pi_r$ the HS operator $\cI(g)$ is self-adjoint.
The maps~(\ref{triv}) and (\ref{refl}) just considered do not involve $J_N$. But it is clear that even when we start from a~self-dual $g$, the maps (\ref{triv}) give rise to new couplings $\hat{g}$ that are generically not self-dual. When also $\hat{g}\in\Pi_r$, we can act with $J_N$ to obtain $\hat{g}'\in\Pi_r$ for which $H(\hat{g}')$ has the same spectrum as $H(g)$ (by virtue of Theorem~\ref{theorem2.6}).

All of the maps studied thus far are involutions. The question now arises which group is generated by these involutions. We proceed to answer this question.

For this purpose it is expedient to switch to dif\/ferent couplings. We do this in two steps. First, we set
\begin{gather}\label{lg}
\lambda =g-\zeta/2,\qquad \zeta\equiv (1,1,1,1),
\end{gather}
so that
\begin{gather}\label{Hl}
H(g)\to -\frac{d^2}{dx^2}+\sum_t\left(\lambda_t^2-\frac{1}{4}\right)\wp(x+\omega_t).
\end{gather}
Also, (\ref{triv}) and (\ref{refl}) become
\begin{gather}\label{ltr}
\lambda_2\to -\lambda_2,\qquad \lambda_3\to -\lambda_3,
\\
\label{lre}
(\lambda_0,\lambda_1,\lambda_2,\lambda_3)\to (\lambda_1,\lambda_0,\lambda_3,\lambda_2),
\end{gather}
and we have
\begin{gather*}
\lambda = \lambda' \ \Leftrightarrow  \ \lambda_0 +\lambda_2 =\lambda_1 + \lambda_3,
\end{gather*}
since $\zeta$ is self-dual.

The second step may seem unmotivated at f\/irst sight. It consists in def\/ining couplings
\begin{gather}\label{cl}
c_0=\lambda_0 +\lambda_3,\qquad c_3=\lambda_0 -\lambda_3, \qquad c_1=\lambda_1 +\lambda_2,\qquad c_2=\lambda_1 -\lambda_2.
\end{gather}
The point of this def\/inition is that the maps (\ref{ltr}) become permutations
\begin{gather}\label{ctr}
c_0\leftrightarrow c_3,\qquad c_1\leftrightarrow c_2,
\end{gather}
while (\ref{lre}) turns into the permutation
\begin{gather}\label{cre}
(c_0,c_1,c_2,c_3)\to (c_1,c_0,c_3,c_2).
\end{gather}
Moreover, the $J_N$-action transforms into the permutation
\begin{gather}\label{Jc}
c_2\leftrightarrow c_3,
\end{gather}
as is easily checked.

Since the three transpositions (\ref{ctr}) and (\ref{Jc}) already generate $S_4$, it follows that (\ref{cre}) may be viewed as a suitable product of these maps. The symmetries (\ref{ctr}), (\ref{cre}) generate an 8-element $S_4$-subgroup isomorphic to the dihedral group~$I_2(4)$. This subgroup gives rise to a~decomposition of~$S_4$ into three cosets. Specif\/ically, as coset-representants we may choose the three permutations
\begin{gather}\label{3per}
(c_0,c_1,c_2,c_3)\ \to \ (c_0,c_1,c_2,c_3),\   (c_0,c_1,c_3,c_2),\  (c_0,c_2,c_3,c_1).
\end{gather}

Now that we have clarif\/ied the character of the symmetry group by exploiting the new couplings $c$, we may and will transform back to $g$. In particular, for  a given $g$ the three maps~(\ref{3per}) are
\begin{gather*}
g\to g,g',\tilde{g},
\end{gather*}
where
\begin{gather*}  \tilde{g}_0\equiv \frac{1}{2}(g_0+g_1+g_2+g_3-1),\qquad \tilde{g}_1\equiv \frac{1}{2}(g_0+g_1-g_2-g_3+1),
  \nonumber\\
 \tilde{g}_2\equiv \frac{1}{2}(-g_0+g_1-g_2+g_3+1),\qquad \tilde{g}_3\equiv \frac{1}{2}(g_0-g_1-g_2+g_3+1).
\end{gather*}
We now summarize and extend the above analysis in the following theorem.

\begin{theor} \label{theorem3.1} The group $G$ generated by the maps
\begin{gather}\label{g1}
(g_0,g_1,g_2,g_3)   \to (g_0,g_1,1-g_2,g_3),
\\
\label{g2}
(g_0,g_1,g_2,g_3)   \to (g_0,g_1,g_2,1-g_3),
\end{gather}
and
\begin{gather}\label{g3}
g\to J_Ng,
\end{gather}
with $J_N$ given by \eqref{defJN}, is isomorphic to $S_4$ and contains the mirror map
\begin{gather*}
(g_0,g_1,g_2,g_3)   \to (g_1,g_0,g_3,g_2).
\end{gather*}
Now assume $g$ belongs to the set $\Pi_G$, cf.~\eqref{PiG}.
Then the $G$-orbit of $g$ belongs to $\Pi_G$ and one has
\begin{gather}\label{sprel}
\sigma (H(g))= \sigma(H(w(g))),\qquad \forall \, w\in G.
\end{gather}
\end{theor}

\begin{proof} We have already shown the f\/irst assertion. Next, we recall that $\Pi_r$ is def\/ined by the f\/ive linear constraints
\begin{gather*}
2g_0>-1,\qquad 2g_1>-1,\qquad g_0+g_1+g_2+g_3>0,\qquad g_0+g_1-g_2+g_3>-1,\\
 g_0+g_1+g_2-g_3>-1.
\end{gather*}
Together with the extra constraint def\/ining $\Pi_G$, i.e.,
\begin{gather*}
g_0+g_1-g_2-g_3>-2,
\end{gather*}
this yields six constraints that transform into
\begin{gather}\label{six}
c_{\mu}+c_{\nu}>-2,\qquad \mu,\nu\in \{ 0,1,2,3\},\qquad \mu\ne\nu,
\end{gather}
as is readily verif\/ied from (\ref{cl}) and (\ref{lg}). Since the latter are manifestly permutation invariant, the $G$-orbit assertion follows. Finally, since (\ref{sprel}) holds for the generators (\ref{g1})--(\ref{g3}) and $\Pi_G$ is invariant under $G$, it holds for all $w\in G$.
\end{proof}

Rewriting (\ref{Hl}) in terms of $c$, we obtain
\begin{gather*}
\hat{H}(c)       =       -\frac{d^2}{dx^2}-\frac{1}{4}\sum_{t=0}^3\wp(x+\omega_t)
 +\frac{1}{4}\left(c_0^2+c_3^2\right)\sum_{t=0,3}\wp(x+\omega_t)+\frac{1}{2}c_0c_3(\wp(x)-\wp(x+\omega_3))
\nonumber\\
\phantom{\hat{H}(c)  = }{}  +\frac{1}{4}\left(c_1^2+c_2^2\right)\sum_{t=1,2}\wp(x+\omega_t)+\frac{1}{2}c_1c_2(\wp(x+\omega_1)-\wp(x+\omega_2)).
   \end{gather*}
Requiring (\ref{six}), the spectral invariance under the $I_2(4)$ subgroup is  readily understood in this guise as well.  But the invariance under any other $c$-permutation looks bizarre.

\section{Special cases}\label{section4}

It is plain from (\ref{Hg}), (\ref{Vg}) that the Heun dif\/ferential operator reduces to $-d^2/dx^2$ for the sixteen~$g$'s obtained by choosing $g_t\in \{ 0,1\}$, $t=0,1,2,3$. Viewing $-d^2/dx^2$ as an operator on~$\cH$ with domain~$\dom$, it is also clear which ONB $\{ b_n\}_{n\in\N}$ of eigenvectors in $\dom$ arises. Specif\/ically, with the normalization constant
\begin{gather*}
\cN \equiv (4r/\pi)^{1/2},
\end{gather*}
the following ONBs are involved:
\begin{gather*}
{\rm (i)}\ \ g_0=g_1=0\ \Rightarrow\ b_0=\cN /\sqrt{2},\ \ \ b_n=\cN \cos 2nrx,\ \ \ n>0,
\\
{\rm (ii)}\ \ g_0=g_1=1\ \Rightarrow\ b_n=\cN \sin (2n+2)rx,
\\
{\rm (iii)}\ \ g_0=1,\ g_1=0\ \Rightarrow\ b_n=\cN \sin (2n+1)rx,
\\
{\rm (iv)}\ \ g_0=0,\ g_1=1\ \Rightarrow\ b_n=\cN \cos (2n+1)rx.
\end{gather*}

We proceed to study the above $g$-choices in relation to $\cI(g)$. For case (i) we must choose $g_2=g_3=1$ to obtain $g\in\Pi_r$. (For $g_2=g_3=0$ we get $s_g=0$, and for $g_2=0$, $g_3=1$ and $g_2=1$, $g_3=0$ we get $g_1'=-1/2$ and $g_0'=-1/2$, resp.) Then we obtain from (\ref{Psi})--(\ref{cS}) and~(\ref{w2})
\begin{gather*}
\Psi((0,0,1,1);x,y)=R(x)R(\omega_1-x)R(y)R(\omega_1-y)/R(x+y)R(x-y).
\end{gather*}
It now follows from Theorem~\ref{theorem2.4} that  the ONB (i) consists of eigenvectors for $\cI((0,0,1,1))$. In fact, this can be readily verif\/ied  by a contour integration, which also yields the eigenvalues explicitly. (To simplify the residues, one needs the duplication formula for the $R$-function, cf.~(\ref{dupl}).) Specif\/ically, we obtain
\begin{gather}\label{nun}
\nu_n=\pi/p\cosh (nr\alpha),\qquad g=(0,0,1,1),\qquad c=(0,0,-1,-1),
\end{gather}
with $p$ the inf\/inite product in (\ref{Rs}).

Turning to case (ii), we get two self-dual subcases. For the f\/irst one,
\begin{gather*}
\Psi((1,1,0,0);x,y)=p^4 s(x)s(\omega_1-x)s(y)s(\omega_1-y)/R(x+y)R(x-y),
\end{gather*}
the eigenvalues of the corresponding integral operator $\cI(g)$ on the ONB (ii) can be calculated in the same way as before. The result is
\begin{gather*}
\nu_n=\pi e^{r\alpha}/p\cosh ((n+1) r\alpha),\qquad g=(1,1,0,0),\qquad c=(0,0,1,1).
\end{gather*}
Using (\ref{dupl}) and (\ref{Rs}), the second subcase can be rewritten as
\begin{gather*}
\Psi((1,1,1,1);x,y)=p^2s(2x)s(2y)/R^2(x+y)R^2(x-y).
\end{gather*}
With due ef\/fort (involving in particular a suitable use of (\ref{Rs})), the eigenvalues for this $g$-choice can again be determined explicitly, yielding
\begin{gather*}
\nu_n=\frac{2\pi(n+1)re^{r\alpha}}{p^2\sinh ((n+1)r\alpha)},\qquad g=(1,1,1,1),\qquad c=(1,1,0,0).
\end{gather*}
It should be noted that the $\cI(g)$-spectra of these two subcases do not coincide, even though they are related by a $c$-permutation.

The remaining two subcases are not self-dual, but they are related by mirror symmetry, cf.~(\ref{mir}). Hence we only study
\begin{gather}\label{remc}
g=(1,1,0,1),\ c=(1,0,1,0) \ \Leftrightarrow \ g'=\frac{1}{2}(3,1,1,1),\ c'=(1,0,0,1).
\end{gather}
For this choice it is clear that $H(g')$ has a nontrivial potential. Its eigenvector ONB is related to the ONB (ii) via $\cI(g')$. Specif\/ically, we must have
\begin{gather*}
b_n(g';x)= \cN_n \intI dy\frac{[w(g';x)w(g;y)]^{1/2}}{[R(x+y)R(x-y)]^{3/2}}\sin (2n+2)ry,\qquad g=(1,1,0,1),
\end{gather*}
where the normalization constant $\cN_n$ ensures that $b_n(g';x)$ has norm 1. Also, the $H(g')$-eigenvalues are given by
\begin{gather}\label{En1}
E_n((3,1,1,1)/2)=(2n+2)^2r^2.
\end{gather}

The cases (iii) and (iv) are related by mirror symmetry, so we only consider case (iii). First we choose
\begin{gather*}
g=(1,0,1,1),\ c=(1,0,-1,0)\ \Leftrightarrow \ g'=\frac{1}{2}(1,1,1,3),\ c'=(1,0,0,-1).
\end{gather*}
Thus $H(g')$ has a nontrivial potential. Just as for (\ref{remc}), its eigenvector ONB is given by
\begin{gather*}
b_n(g';x)= \cN_n \intI dy\frac{[w(g';x)w(g;y)]^{1/2}}{[R(x+y)R(x-y)]^{3/2}}\sin (2n+1)ry,\qquad g=(1,0,1,1).
\end{gather*}
 The analog of (\ref{En1}) is
\begin{gather*}
E_n((1,1,1,3)/2)=(2n+1)^2r^2.
\end{gather*}

We continue with
\begin{gather*}
g=(1,0,0,0),\ c=(0,-1,0,1) \ \Leftrightarrow \ g'=\frac{1}{2}(1,1,-1,1),\ c'=(0,-1,1,0).
\end{gather*}
Here we obtain the $H(g')$-ONB
\begin{gather*}
b_n(g';x)=\cN_n  \intI dy\frac{[w(g';x)w(g;y)]^{1/2}}{[R(x+y)R(x-y)]^{1/2}}\sin (2n+1)ry,\qquad g=(1,0,0,0),
\end{gather*}
with  eigenvalues
\begin{gather*}
E_n((1,1,-1,1)/2)=(2n+1)^2r^2.
\end{gather*}
Next we consider the self-dual choice $g_2=0$, $g_3=1$. Then we get
\begin{gather*}
\Psi((1,0,0,1);x,y)=p^2s(x)R(\omega_1-x)s(y)R(\omega_1-y)/R(x+y)R(x-y).
\end{gather*}
As before, the eigenvalues of $\cI(g)$ on the ONB (iii) can be explicitly determined, yielding
\begin{gather*}
\nu_n=\pi e^{r\alpha/2}/p\cosh ((n+1/2)r\alpha),\qquad g=(1,0,0,1),\qquad c=(1,-1,0,0).
\end{gather*}

Finally, we study the choice
\begin{gather*}
g=(1,0,1,0),\ c=(0,0,-1,1) \ \Leftrightarrow \ g'=(0,1,0,1),\ c'=(0,0,1,-1),
\end{gather*}
for which
\begin{gather*}
\Psi((1,0,1,0);x,y)=p^2s(x)R(x)s(\omega_1-y)R(\omega_1-y)/R(x+y)R(x-y).
\end{gather*}
Once more, a contour integration yields the expected result
\begin{gather*}
\sin (2n+1)rx =
\nu_n  \intI dy \Psi((1,0,1,0);x,y)\cos (2n+1)ry,
\end{gather*}
with the singular values $\nu_n$ given by
\begin{gather*}
\nu_n = \pi e^{r\alpha/2}/p\sinh((n+1/2)r\alpha),\qquad g=(1,0,1,0),\qquad c=(0,0,-1,1).
\end{gather*}

We conclude this section with some remarks on special cases of a dif\/ferent character. We recall f\/irst that $\cI(g)$ reduces to a rank-one operator for $g\in\Pi$ with $s_g=0$, cf.~(\ref{Psi}) and~(\ref{cS}). By continuity in $g$, it therefore follows that the functions $w(g;x)^{1/2}$ and $w(g';x)^{1/2}$ are eigenfunctions of $H(g)$ and $H(g')$, resp., with the same eigenvalue. We point out that this is not at all obvious. With due ef\/fort, however, this can be directly verif\/ied.

Specif\/ically, it follows from I (3.79), (3.113) and (3.114) that the eigenfunction property is equivalent to constancy of the function
\begin{gather*}
\sum_{t=0}^3g_t^2\wp(x+\omega_t)-\left(\sum_{t=0}^1\left( g_t\frac{s'(x+\omega_t)}{s(x+\omega_t)}+
g_{t+2}\frac{R'(x+\omega_t)}{R(x+\omega_t)}\right)\right)^2,\qquad \sum_{t=0}^3g_t=0.
\end{gather*}
We encountered this functional identity before on p.~253 of~\cite{cadiz}, where we supplied a (short) proof. Using next I (3.105)--(3.106), the constant (which amounts to the ground state eigenvalue) can be calculated explicitly:
\begin{gather*}
E_0(g)=\sum_{j=1}^3g_je_j(g_j+2g_0),\qquad e_j\equiv \wp(\omega_j),\qquad j=1,2,3, \qquad \sum_{t=0}^3g_t=0.
\end{gather*}
From this formula one readily verif\/ies equality to $E_0(g')$.

Finally, we comment on the case where $V(g;x)$ is real-analytic on $\R$. For this we need to choose $g_0,g_1\in\{ 0,1\} $,  cf.~(\ref{Vg}). Then one need only combine the $\cH$-ONB's $b_n((0,0,g_2,g_3);x)$ and $b_n((1,1,g_2,g_3);x)$ (consisting of real-analytic $\pi/r$-periodic functions that are even and odd, resp.) to obtain an ONB of $\pi/r$-periodic functions for
\begin{gather*}
\cH_b\equiv L^2([-\pi/2r,\pi/2r],dx).
\end{gather*}
Likewise, an ONB of $\pi/r$-antiperiodic functions for $\cH_b$ results from combining the $\cH$-ONB $b_n((0,1,g_2,g_3);x)$ of even functions and the $\cH$-ONB of odd functions $b_n((1,0,g_2,g_3);x)$. In terms of Floquet theory, therefore, one can only arrive at the $\cH_b$-ONB's associated with the multipliers~1 and $-1$. Whenever the four $g$-values are all in $\Pi_r$, one can invoke the spectral invariance under taking $g\to g'$. Note that $V(g';x)$ may be singular at $x=0$ and/or $x=\pi/2r$, as we have already seen in the above special cases where $V(g;x)$ vanishes identically.

\appendix
																																												 
\section{Proofs of Lemmas \ref{lemma2.1} and \ref{lemma2.2}}\label{appendix-A}

\begin{proof}[Proof of Lemma \ref{lemma2.1}] We recall that the function
\begin{gather*}
h(x)\equiv \intI dy\cS(g;x,y)\phi(y),\qquad \phi \in L^1([0,\pi/2r]),\qquad x\in [0,\pi/2r],
\end{gather*}
extends to an even analytic function in $|\im z|<\alpha/2$, cf.~the paragraph containing (\ref{hdef}). Thus its vanishing is equivalent to vanishing of the coef\/f\/icients in its Taylor expansion
\begin{gather}\label{hcn}
h(x)=\sum_{k=0}^{\infty}c_kx^{2k},
\end{gather}
at $x=0$. We proceed to study these coef\/f\/icients.

Obviously, we need  information on the coef\/f\/icients in the expansion of
$\cS(g;x,y)$ at $x=0$. To this end we f\/irst consider the expansion
\begin{gather*}
F(x)    \equiv   -\frac{1}{2}( \ln R(x+y)+\ln R(x-y))
    =     \sum_{j=0}^{\infty}\frac{F^{(2j)}(0)}{(2j)!}x^{2j}.
  \end{gather*}
  Using
  \begin{gather*}
  -\frac{d^2}{dz^2}\ln R(z)=\tP(z)+2\eta r/\pi,\qquad \tP(z)\equiv \wp(z+i\alpha/2),
  \end{gather*}
  (cf.~(\ref{Rs}), (\ref{ssig})), we obtain
  \begin{gather*}
  F(0)=-\ln R(y),\qquad F^{(2)}(0)=\tP(y)+2\eta r/\pi,
\\
  F^{(2k)}(0)=\tP^{(2k-2)}(y),\qquad k>1.
  \end{gather*}
  Now from standard elliptic function lore~\cite{ww} we deduce
  \begin{gather*}
  \wp^{(2k-2)}(z)=(2k-1)!\wp(z)^k +  {\rm l.d.},\qquad k>1,
  \end{gather*}
  where l.d.\ stands for a polynomial in $\wp(z)$ of degree $<k$. Thus we have
  \begin{gather*}
  F(x)=-\ln R(y)+\sum_{j=1}^{\infty}a_j x^{2j},
  \end{gather*}
  with
  \begin{gather}\label{ak}
  a_k=\frac{1}{2k}\tP(y)^k +  {\rm l.d.}
  \end{gather}

  We are now prepared to consider the expansion of $\cS$. It is given by
  \begin{gather}
  \cS(g;x,y)     =     R(y)^{-2s_g}\exp\left( 2s_g\sum_{j=1}^{\infty}a_jx^{2j}\right)
  =    R(y)^{-2s_g}\sum_{k=0}^{\infty}b_kx^{2k},\label{Sex}
\end{gather}
   so that the f\/irst few $b_k$ read
   \begin{gather*}
   b_0=1,\qquad b_1=2s_ga_1,\qquad b_2=2s_ga_2+2s_g^2a_1^2,\qquad b_3=2s_ga_3+4s_g^2 a_2a_1 + 4s_g^3 a_1^3/3.
   \end{gather*}
 More generally, it is easily seen that $b_k$ is the solution to the recurrence
   \begin{gather*}
   kb_k=2s_g\sum_{j=1}^k ja_j b_{k-j},\qquad  k>0,\qquad b_0=1.
   \end{gather*}
 In view of (\ref{ak}),  this implies that $b_k$ is given by
 \begin{gather}\label{bQ}
 b_k=Q^{(k)}(s_g)\tP(y)^k+  {\rm l.d.},
 \end{gather}
 where $Q^{(k)}(u)$ is a polynomial of the form
 \begin{gather*}
 Q^{(k)}(u)=\sum_{l=1}^kp_l^{(k)}u^l,\qquad p^{(k)}_l\in(0,\infty).
 \end{gather*}

 We are now in the position to use our assumption $s_g>0$. It entails that the coef\/f\/i\-cient~$Q^{(k)}(s_g)$ in (\ref{bQ}) is positive:
 \begin{gather}\label{bkp}
 b_k=q_k\tP(y)^k+ {\rm l.d.},\qquad q_k\in(0,\infty),\qquad \forall \, k\in\N.
 \end{gather}
 Returning to (\ref{hcn}), we f\/irst note that (\ref{Sex}) entails
 \begin{gather*}
 c_k=\intI dyb_k(y)\psi(y),
 \end{gather*}
 where
 \begin{gather}\label{psiR}
 \psi(y)\equiv R(y)^{-2s_g}\phi(y).
 \end{gather}
 Since the coef\/f\/icients $c_k$ vanish by assumption, it follows recursively from (\ref{bkp}) that we have
 \begin{gather*}
 \intI dy \tP(y)^k\psi(y)=0,\qquad \forall \, k\in\N.
 \end{gather*}

 Now $\tP (y)$ is monotone increasing on $[0,\pi/2r]$. Thus we can invoke the Stone-Weierstrass theorem~\cite{rs1} to conclude that the span of the functions $1,\tP(y),\tP(y)^2,\tP(y)^3,\ldots$ is dense (in the supremum norm) in the space of real-valued continuous functions on $[0,\pi/2r]$. Since $\psi(y)$ is an $L^1$-function, it follows by dominated convergence that we have
 \begin{gather*}
 \intI dy C(y)\psi(y)=0,\qquad \forall \, C\in C_{\R}([0,\pi/2r]).
 \end{gather*}
 Hence $\psi=0$, so by (\ref{psiR}) $\phi=0$.
\end{proof}

\begin{proof}[Proof of Lemma~\ref{lemma2.2}] Letting $e(x)\in \dom$, we f\/irst show that the action of the dif\/ferential operator $H(g)$ (\ref{Hg}) on $e(x)$ yields a function in $\cH$. To this end we use the relations
 \begin{gather*}
 e(x)=s(x)^{g_0}s(x-\omega_1)^{g_1}d(x),\qquad d\in\cD,
 \end{gather*}
 and
 \begin{gather*}
 \frac{s'(z)^2}{s(z)^2}=\frac{s''(z)}{s(z)}+\wp(z)+\eta \omega_1,
 \end{gather*}
 to calculate
 \begin{gather*}
e''(x)     =    e(x)\sum_{t=0,1}\left(
[g_t^2-g_t][\wp(x+\omega_t)+\eta \omega_1]
+g_t^2\frac{s''(x+\omega_t)}{s(x+\omega_t)}
\right)
\nonumber\\
\phantom{e''(x)     =}{}  +e(x)\left(
  Q(x)+2\frac{d'(x)}{d(x)}\sum_{t=0,1}g_t\frac{s'(x+\omega_t)}{s(x+\omega_t)}
   +\frac{d''(x)}{d(x)}\right),
  \end{gather*}
where
\begin{gather}\label{defQ}
Q(x)\equiv 2g_0g_1 \frac{s'(x)}{s(x)} \frac{s'(x+\omega_1)}{s(x+\omega_1)}.
\end{gather}
Subtracting the potential term
\begin{gather*}
e(x)\sum_{t=0}^3 g_t(g_t-1)\wp(x+\omega_t),
\end{gather*}
we obtain a sum of terms that are manifestly in $\cH$, save for the terms
\begin{gather}\label{rem}
e(x)\left( Q(x)+\sum_{t=0,1} g_t^2\frac{s''(x+\omega_t)}{s(x+\omega_t)}\right)
+2s(x)^{g_0}s(x-\omega_1)^{g_1}
\sum_{t=0,1}g_td'(x)\frac{s'(x+\omega_t)}{s(x+\omega_t)}.
\end{gather}
Now $s(z)$ is odd and has a f\/irst order zero for $z=0$, so $s''(z)$ vanishes at $z=0$ at least to f\/irst order. Thus the two terms in the f\/irst sum are smooth for $x\in\R$.
Next, $s'(z)$ vanishes at $z=\omega_1$, since $s(z)$ is odd and $2\omega_1$-antiperiodic. Therefore $Q(x)$~(\ref{defQ}) is also smooth. Finally, since $d'(0)=d'(\omega_1)=0$, the two  terms in the second sum are continuous on $[0,\omega_1]$. Hence the function given by (\ref{rem}) is in $\cH$, so that $H(g)$ is well def\/ined on $\dom$.

It remains to show that $H(g)$ is symmetric on $\dom$. To this end, let $d_1,d_2\in\cD$ and consider
\begin{gather}
(s(\cdot)^{g_0}s(\cdot +\omega_1)^{g_1}d_1,H(g)s(\cdot)^{g_0}s(\cdot +\omega_1)^{g_1}d_2)\nonumber\\
\qquad{}-
(H(g)s(\cdot)^{g_0}s(\cdot +\omega_1)^{g_1}d_1,s(\cdot)^{g_0}s(\cdot +\omega_1)^{g_1}d_2).\label{ipd}
\end{gather}
From the above calculation we see that this equals
\begin{gather*}
  \int_0^{\omega_1} dx s(x)^{2g_0}s(x+\omega_1)^{2g_1}\biggl( \overline{d_1^{''}(x)}d_2(x)-\overline{d_1(x)}
d_2^{''}(x)
   \nonumber\\
\qquad{} +2[\overline{d_1^{'}(x)}d_2(x)-
\overline{d_1(x)}d_2^{'}(x)]\sum_{t=0,1}g_t\frac{s'(x+\omega_t)}{s(x+\omega_t)}\biggr).
\end{gather*}
Clearly, this can be rewritten as
\begin{gather}\label{dd}
  \int_0^{\omega_1}\! dx \left(\!-\overline{d_1(x)}\frac{d}{dx}[s(x)^{2g_0}s(x+\omega_1)^{2g_1}d_2'(x)]
  +d_2(x)\frac{d}{dx}[s(x)^{2g_0}s(x+\omega_1)^{2g_1}\overline{d_1'(x)}]\!\right)\!.\!\!\!
  \end{gather}
Now for all $d\in\cD$ we have
\begin{gather*}
d'(x)=\int_0^x d''(t)dt =O(x),\quad x\downarrow 0,\qquad d'(x)=-\int_x^{\omega_1}d''(t)dt=O(x-\omega_1),\quad x\uparrow \omega_1.
\end{gather*}
Integrating by parts in (\ref{dd}), it follows that the boundary terms vanish, whereas the new integrand vanishes identically. As a result, the dif\/ference (\ref{ipd}) vanishes, so that $H(g)$ is indeed symmetric on~$\dom$.
\end{proof}

\subsection*{Acknowledgments}

We would like to thank F.~Nijhof\/f, B.~Sleeman and K.~Takemura for illuminating discussions and for supplying information about related literature.

\pdfbookmark[1]{References}{ref}
\LastPageEnding

\end{document}